\documentclass[twocolumn,prd]{revtex4}%
\usepackage{graphicx}
\usepackage{amsmath}
\usepackage{amsfonts}
\usepackage{amssymb}%
\newcommand{\f}{\begin{equation}}
\newcommand{\ff}{\end{equation}}
\newcommand{\fa}{\begin{eqnarray}}
\newcommand{\ffa}{\end{eqnarray}}

\begin{document}
\title{The kinematics of particles moving in rainbow spacetime}
\author{Yi Ling$^{1,2}$}\email{yling@ncu.edu.cn}
\affiliation{${}^1$ Center for Gravity and Relativistic
Astrophysics, Nanchang University, 330047, China}
\affiliation{%
${}^2$ CCAST (World Laboratory), P.O. Box 8730, Beijing
   100080, China}
\author{Song He}
 \affiliation{Institute of Theoretical Physics, School of Physics, Peking
University, Beijing, 100871, China}

\author{Hongbao Zhang}
    \affiliation{Department of Astronomy, Beijing Normal University, Beijing, 100875,
    China\\
    Department of Physics,
Beijing Normal University, Beijing, 100875, China\\
 CCAST (World
Laboratory), P.O. Box 8730, Beijing,
   100080, China}
\begin{abstract}
The kinematics of particles moving in rainbow spacetime is studied
in this paper. In particular the geodesics of a massive particle in
rainbow flat spacetime is obtained when the semi-classical effect of
its own energy on the background is taken into account. We show that
in general the trajectory of a freely falling particle remains
unchanged which is still a straight line as in the flat spacetime.
The implication to the Unruh effect in rainbow flat spacetime is
also discussed.
\end{abstract}

 \maketitle

\section{Introduction}
Recently the formalism of rainbow gravity has been proposed as a
generalization of doubly special relativity when incorporating with
curved spacetime\cite{Magueijo02xx}. It can be viewed as a
phenomenological model at the semi-classical level of quantum
gravity where the quantum effect of moving particles on the
background is taken into account. One key ingredient in this
formalism is that there is no single fixed background for all
observers, but dependent on the energy $E$ of probes. Corresponding
to a modified dispersion relation as proposed in doubly special
relativity\cite{Amelino00ge,Amelino00mn,Amelino03ex,Amelino03uc,Magueijo01cr,Magueijo02am,Smolin05cz},
the dual or position space is defined by requiring that the
contraction between momentum and infinitesimal displacement be a
linear invariant. Specifically, given a modified dispersion relation
as \f E^2f_1^2(E,\eta)-p^2f_2^2(E,\eta)=m_0^2,\ff the dual space
$dx^a$ is endowed with an energy dependent invariant \f
ds^2=-{1\over f_1^2(E,\eta)}dt^2+{1\over
f_2^2(E,\eta)}dx^2,\label{rfs}\ff where $f_1$ and $f_2$ are two
general functions of energy $E$ and $\eta$ which is a dimensionless
parameter labeling the magnitude of correction terms. To go back to
special relativity as $E/E_{p}\ll 1$ where $E_p$ is the Planck
energy, one requires that $f_1(E,\eta)$ and $f_2(E,\eta)$ approach
to one at this limit\footnote{In \cite{Magueijo02xx} the coordinates
$t,x$ are also named as energy independent ones. Alternatively some
energy dependent coordinates can be defined. The relations between
these two sorts of coordinates and their physical interpretations
can be found in \cite{Magueijo02xx} as well.}.

The invariant in equation (\ref{rfs}) is usually named as a rainbow
flat metric. It can be extended to incorporate with curvature when
the usual effect of classical gravity is taken into account. In this
general case the background spacetime is described by a rainbow
metric which is given in terms of a one-parameter family of
orthonormal frame fields, \f g(E)=\eta^{ab}e_a(E)\otimes e_b(E).\ff
Then through the standard process, the corresponding one-parameter
family of connections $\nabla(E)_{\mu}$ and curvature tensors
$R(E)^\rho_{\mu\nu\lambda}$ can be constructed, leading to modified
Einstein's field equations \f G_{\mu\nu}(E)=8\pi
G(E)T_{\mu\nu}(E)+g_{\mu\nu}(E)\Lambda (E),\label{MEE}\ff where
Newton's constant as well as the cosmological constant is
conjectured to be energy dependent as one expects from the viewpoint
of renormalization group theory, for instance $G(E)=g^2(E)G$ and
$\Lambda (E)=h^2(E)\Lambda$.

This proposal has received considerable attention recently and other
stimulated work on this formalism can be found in
\cite{Galan04st,Galan05ju,Hackett05mb,Aloisio05qt,Ling05bp,Galan06by,Ling06}.
In this paper we intend to study the kinematics of massive particles
moving in such a one-parameter spacetime described by rainbow
metric. In particular the semi-classical effect on the background
due to its own energy of the moving particle is concerned. We
firstly present a general discussion on the equation of geodesics of
particles moving in rainbow spacetime, then consider solutions to
this equation in the special case that the background is described
by the rainbow flat metric. Interestingly, our results show that in
this case the geodesics is the same as that one in usual flat
spacetime. We will briefly discuss the possible implications to the
Unruh effect as well.

\section{Particles moving in rainbow spacetime}
For dust in rainbow spacetime, its energy momentum tensor can be
written as \f T_{\mu\nu}(E)=\rho T_\mu(E)T_\nu(E),\ff where
$T_\mu(E)$ is the four-velocity of the particle, satisfying the
ordinary normalized condition\f
T^\mu(E)T_\mu(E)=-1.\label{normalized}\ff Now using the Bianchi
identity or the conservation of the energy momentum tensor, we have
\fa &&\rho\nabla (E)_{\mu}T^{\mu}(E)T^\nu(E)+\rho T^\mu(E)\nabla
(E)_{\mu}T^\nu(E)\nonumber\\
&&+T^\mu(E)T^\nu(E)\nabla (E)_\mu\rho=0.\ffa Furthermore,
contracting both sides with ${\delta_\nu}^\lambda+T_\nu T^\lambda$
and employing the normalized condition (\ref{normalized}), the above
equation yields \f T^\mu\nabla (E)_\mu T^\lambda=0,\label{gg}\ff
which shows that a free particle also goes along the geodesics in
rainbow spaceime.

\section{Geodesics in rainbow flat spacetime}
In this section, we shall investigate the geodesics of a single
particle with energy $E$ moving in rainbow flat spacetime which is
described by the metric (\ref{rfs}). Particularly we intend to
answer the question of how its trajectory may be changed when the
semi-classical effect due to its own energy on the background is
taken into account. In this case the energy appearing in the rainbow
metric $(\ref{rfs})$ is identified with the energy of the particle
itself. Thus recalling the general definition of the energy of a
massive particle in curved spacetime, we have \f
E=-g_{00}P^0={m_0\over f_1^2(E)}{dt\over ds},\label{ep} \ff where
$s$ is understood as the proper time of this particle, and for
convenience the parameter $\eta$ in the function $f_1$ has been set
as unit since it does not play an important role in our present
discussion. On the other hand from (\ref{gg}) we may write the
equation of geodesics as \f {d^2x^{\mu}\over
ds^2}+\Gamma^{\mu}_{\rho\sigma}(E){dx^{\rho}\over
ds}{dx^{\sigma}\over ds}=0.\label{gop}\ff Obviously if $
f_1^2(E)=f_2^2(E)=1,$ we have $\Gamma^{\mu}_{\rho\sigma}(E)=0$ and
the corresponding geodesics is a straight line in Minkowski
spacetime. But in general, it is unclear whether the connection
still vanishes since it manifestly depends on the energy of the
moving particle, which should not be thought of as a constant for
granted. As a matter of fact, given a specific form of the function
$f_1$ we may obtain the energy of the particle in terms of the
zeroth component of the four-velocity, namely ${dt\over ds}$, from
(\ref{ep}), while whether the quantity of ${dt\over ds}$ is constant
or not is completely determined by the equation of geodesics.
Therefore to obtain the trajectory of a freely falling particle
moving in rainbow spacetime we need solve for both the geodesics
equation (\ref{gop}) and the energy equation (\ref{ep}) together.

Next we will show that the energy of the moving particle remains a
constant indeed by solving these two equations. For simplicity but
without loss of generality, we consider the case in two dimensional
spacetime, where the components of connection are: \fa
 \Gamma^{0}_{00}&&=-{1\over
f_1}{\partial f_1\over \partial t},\ \ \
\Gamma^{0}_{11}=-{f_1^2\over f_2^3}{\partial f_2\over
\partial t},\nonumber\\
\Gamma^{0}_{01}&&=\Gamma^{0}_{10}=-{1\over
f_1}{\partial f_1\over \partial x},\nonumber\\
\Gamma^{1}_{00}&&=-{f_2^2\over f_1^3}{\partial f_1\over
\partial x},\ \ \
\Gamma^{1}_{11}=-{1\over f_2} {\partial f_2\over
\partial x},\nonumber\\
\Gamma^{1}_{01}&&=\Gamma^{1}_{10}=-{1\over f_2}{\partial f_2\over
\partial t}.\ffa Thus the equations of geodesics reads
\fa&&{d^2t\over ds^2}-{1\over f_1}{\partial f_1\over \partial t}
\left({dt\over ds}\right)^2-{f_1^2\over f_2^3}{\partial f_2\over
\partial t}\left({dx\over ds}\right)^2\nonumber\\
&&-2{1\over f_1}{\partial f_1\over \partial x}{dt\over ds}{dx\over
ds}=0,\nonumber\\
&&{d^2x\over ds^2}-{f_2^2\over f_1^3}{\partial f_1\over
\partial x}\left({dt\over ds}\right)^2-{1\over f_2} {\partial f_2\over
\partial x}\left({dx\over ds}\right)^2\nonumber\\
&&-2{1\over f_2}{\partial f_2\over\partial t}{dt\over ds}{dx \over
ds}=0. \label{geodesics}\ffa  From now on we denote ${dt\over ds}$
as $\lambda$, then from (\ref{ep}) we may express $E$ as a function
of $\lambda$ such that $f_1(E)$ as well as $f_2(E)$ can be rewritten
as \f f_1=G_1(\lambda),\ \ \ f_2=G_2(\lambda).\ff Denoting $\{t,x\}$
as $\{x_0,x_1\}$, respectively, we obtain \f {\partial
f_i\over\partial x_j}={dG_i\over d\lambda}{d\lambda\over
ds}{\partial s\over \partial x_j}={dG_i\over d\lambda}{\partial
s\over \partial x_j}{d^2t\over ds^2},\label{ddd}\ff where $i=1,2$
and $j=0,1$. Thus the first equation in (\ref{geodesics}) reduces
into \fa A{d^2t\over ds^2}=0,\ffa where $A$ is some complicated
coefficient function depending on the specific form of functions
$f_1(E)$ and $f_2(E)$. Usually $A$ does not vanish and we may see
this from the example given below. Hence we finally obtain \fa
{d^2t\over ds^2}=0.\ffa Plugging it into (\ref{ep}) we show that the
energy $E$ of the particle is indeed a constant along the geodesics
in rainbow flat spacetime. As a consequence we also have \f
{\partial f\over
\partial t}={\partial f\over
\partial x}=0,\ff such that all connection components vanish and
the second equation in (\ref{geodesics}) reduces to \f {d^2x\over
ds^2}=0.\ff Therefore the geodesics is still a straight line in
energy-independent coordinate system. Above considerations can be
extended to the four dimensional case straightforwardly as the
equation (\ref{ddd}) plays a crucial role in the proof which is
independent of the dimension of spacetime.

For explicitness we would like to provide an example  by specifying
\f f_1^2(E)=f^2=1-l_pE,\ \ \ \ f_2^2(E)=1,\ff where $l_p\sim 1/E_p$
is the Planck length. Then the non-vanishing components read as \fa
\Gamma^{0}_{00}&=&-{1\over f}{\partial f\over
\partial t},\ \ \ \Gamma^{1}_{00}=-{1\over f^3}{\partial f\over
\partial x},\nonumber\\ \Gamma^{0}_{01}&=& \Gamma^{0}_{10}=-{1\over
f}{\partial f\over
\partial x},\ffa
and the equation of geodesics can be written as
 \fa
&&{d^2t\over ds^2}-{1\over f}{\partial f\over \partial t}
\left({dt\over ds}\right)^2-2{1\over
f}{\partial f\over \partial x}{dt\over ds}{dx\over ds} = 0,\nonumber\\
&&{d^2x\over ds^2}-{1\over f^3}{\partial f\over \partial x}
\left({dt\over ds}\right)^2 = 0.\ffa On the other hand, from
(\ref{ep}) we obtain \f f^4(E)-f^2(E)+l_pm\lambda=0,\ff which yields
a non-perturbative solution to $f$ as \f f^2(E)=
{1+\sqrt{1-4l_pm\lambda}\over 2}\equiv {1+k\over 2}.\label{stf}\ff
This requires
 \f  \lambda\leq {1\over 4l_pm }.\label{cs1}\ff
Thus using (\ref{stf}) the connection components can be computed and
plugging them into the first equation of geodesics, we find it
becomes
 \f
\left({k+3\over 4k}\right){d^2t\over ds^2}=0.\ff  Because of the
constraint (\ref{cs1}), we know ${k+3\over 4k}\neq 0$, thus the
geodesics is the same as the one in flat spacetime ${d^2t\over
ds^2}={d^2x\over ds^2}=0$. Thus $\lambda$ is a constant, and the
energy of the particle has a form \f E={2m\lambda\over
1+\sqrt{1-4l_pm\lambda}}.\ff At the classical limit $l_p\rightarrow
0$, comparing with the standard result in special relativity, we can
easily fix it as \f \lambda={1\over \sqrt{1-v^2}}.\ff The energy has
a cutoff at $\lambda={1\over 4ml_p}$,\f E_{max}={1\over 2l_p},\ff as
one expects from the viewpoint of doubly special relativity.

\section{Implications to Unruh Effect}
In this section we present a brief discussion on Unruh effect in
rainbow flat spacetime.  It is well known that an accelerating
observer in Minkowski spacetime will detect a thermal bath
surrounding him and the Unruh temperature is supposed to be
proportional to the magnitude of the proper acceleration \f T=2\pi a
.\ff This identification has also been justified in de-Sitter and
Anti de-Sitter spacetimes\cite{Deser98bb,Deser98xb}. Here we assume
this identification is still valid in rainbow flat spacetime.
Consider a detector is moving in a spacetime with a rainbow metric
as \f ds^2=-{dt^2\over f^2}+{dx^2\over g^2}, \ff where for
simplicity we take  \f f^2=g^2=1-(l_pE)^2. \label{mdr2} \ff As
usual, we consider the case that the trajectory of the detector in
the energy-independent coordinates is hyperbolic, namely
 \f  x^2-t^2={1\over a^2},  \ff
such that the proper time $\tau$ can be defined as  \f d\tau={1\over
af}d\eta, \ff where \f  x={1\over a} \cosh\eta, \ \ \ \ \ t={1\over
a} \sinh\eta . \ff It is easy to check that the four-velocity
satisfies the normalized condition \f
g_{ab}U^aU^b=g_{ab}\left({\partial \over
\partial \tau} \right)^a\left({\partial \over
\partial \tau} \right)^b=-1.\ff
Furthermore, we may obtain \fa P^0&=&m{dx^0\over
d\tau}=mf(E)\cosh\eta ,\nonumber\\ P^1&=&m{dx^1\over
d\tau}=mf(E)\sinh\eta, \ffa such that the energy and the momentum
read \fa E&=&-g_{00}P^0={m\over f}\cosh\eta \nonumber\\
P&=&g_{11}P^1={m\over f}\sinh\eta.\label{ep2}\ffa Then from
(\ref{mdr2}) and (\ref{ep2}) we have \f
E^2-l_p^2E^4=m^2\cosh^2\eta,\ff such that \f {dE\over
d\eta}=\frac{m^2\cosh\eta \sinh\eta}{E(1-2l_p^2E^2)}.\ff Obviously
along the trajectory the energy of the detector is not a constant
but has a form \f E^2={1-\sqrt{1-4l_p^2m^2\cosh^2\eta}\over
2l_p^2}.\label{eoe}\ff It goes back to the usual form $E=m\cosh\eta$
as $l_p\rightarrow 0$. From (\ref{eoe}) we also notice that the
parameter $\eta$ is constrained by $\cosh\eta\leq {1\over 2ml_p}$,
implying that our classical picture is not valid any more as the
energy of the detector approaches to the Planck scale.

Finally, the proper acceleration for the moving particle can be
computed by \f a^{\mu}={d^2x^{\mu}\over
d\tau^2}+\Gamma^{\mu}_{\rho\sigma}(E){dx^{\rho}\over
d\tau}{dx^{\sigma}\over d\tau},\ff where \fa
\Gamma^{0}_{00}&=&\Gamma^{0}_{11}=\Gamma^{1}_{01}=\Gamma^{1}_{10}=-{1\over
f}{\partial f\over \partial t}\nonumber\\ &=& \frac{am^2l_p^2\sinh\eta}{f^2(1-2l_p^2E^2)},\nonumber\\
\Gamma^{1}_{00}&=&\Gamma^{1}_{11}=\Gamma^{0}_{01}=\Gamma^{0}_{10}=-{1\over
f}{\partial f\over \partial x}\nonumber\\ &=&
\frac{am^2l_p^2\cosh\eta}{f^2(1-2l_p^2E^2)}.\ffa

Then after a long calculation, it turns out that the magnitude of
the acceleration is \fa |a|^2&=&a^2f^2\left\{ \left[1+{l_p^2E^2\over
1-2l_p^2E^2}\left(2-{m^2\over E^2f^2}\right)\right]^2\right.\nonumber\\
 &-&\left.{l_p^4E^4\over (1-2l_p^2E^2)^2}\left(1-{m^2\over
E^2f^2}\right)\right\}. \ffa Obviously this formula can be trusted
only when $E\ll 1/l_p$. Nevertheless, with the assumption that the
thermal effect is proportional to the proper acceleration of the
detector in flat spacetime, this result implies that the usual Unruh
temperature might be modified due to the semi-classical effect of
the detector energy.

\section{Concluding remarks}
In this paper we have attempted to study the kinematics of massive
particles moving in rainbow spacetime. The equation of geodesics has
been given. Especially we considered a single particle moving in
flat rainbow spacetime and proved that its trajectory will not
change due to the semi-classical effect of its own energy. Whether
this conclusion can be extended to other rainbow curved spacetimes
awaits further investigations.

We also considered the possible implications to Unruh effect in
rainbow flat spacetime. Fixing the trajectory of the moving detector
as a hyperbola in energy-independent coordinate system, we find the
proper acceleration receives modifications due to the rainbow effect
of its own energy, implying the Unruh temperature might also be
corrected. It is noteworthy that our discussion presented here is
preliminary and the detailed investigation calls for the quantum
field theory over the curved spacetime endowed with rainbow metrics.

\section*{Acknowledgement}
We are grateful to Prof. Bo Hu and Xiang Li  for helpful
discussions. Y. Ling's work is partly supported by NSFC(No.10405027
) and SRF for ROCS.  S. He's work is supported by NSFC(Nos.10235040
and 10421003). H. Zhang's work is supported in part by
NSFC(Nos.10373003 and 10533010).


\begin{thebibliography}{9}                                                                                                %
%\cite{Magueijo:2002xx}
\bibitem{Magueijo02xx}
  J.~Magueijo and L.~Smolin,
  % ``Gravity's Rainbow,''
  %
  Class.\ Quant.\ Grav.\  {\bf 21}, 1725 (2004)
  [arXiv:gr-qc/0305055].
  %%CITATION = GR-QC 0305055;%%
  %\cite{Amelino-Camelia:2000ge}
\bibitem{Amelino00ge}
  G.~Amelino-Camelia,
   %``Testable scenario for relativity with minimum-length,''
  %
  Phys.\ Lett.\ B {\bf 510}, 255 (2001)
  [arXiv:hep-th/0012238].
  %%CITATION = HEP-TH 0012238;%%

%\cite{Amelino-Camelia:2000mn}
\bibitem{Amelino00mn}
  G.~Amelino-Camelia,
  % ``Relativity in space-times with short-distance structure governed by an
  % observer-independent (Planckian) length scale,''
  %
  Int.\ J.\ Mod.\ Phys.\ D {\bf 11}, 35 (2002)
  [arXiv:gr-qc/0012051].
  %%CITATION = GR-QC 0012051;%%

%\cite{Amelino-Camelia:2003ex}
\bibitem{Amelino03ex}
  G.~Amelino-Camelia, J.~Kowalski-Glikman, G.~Mandanici and A.~Procaccini,
  %``Phenomenology of doubly special relativity,''
  Int.\ J.\ Mod.\ Phys.\ A {\bf 20}, 6007 (2005)
  [arXiv:gr-qc/0312124].
  %%CITATION = GR-QC 0312124;%%

%\cite{Amelino03uc}
\bibitem{Amelino03uc}
  G.~Amelino-Camelia,
   ``The three perspectives on the quantum-gravity problem and their
   implications for the fate of Lorentz symmetry,''
  %
  [arXiv:gr-qc/0309054].
  %%CITATION = GR-QC 0309054;%%

%\cite{Magueijo:2001cr}
\bibitem{Magueijo01cr}
  J.~Magueijo and L.~Smolin,
  %``Lorentz invariance with an invariant energy scale,''
  Phys.\ Rev.\ Lett.\  {\bf 88}, 190403 (2002)
  [arXiv:hep-th/0112090].
  %%CITATION = HEP-TH 0112090;%%

%\cite{Magueijo:2002am}
\bibitem{Magueijo02am}
  J.~Magueijo and L.~Smolin,
  %``Generalized Lorentz invariance with an invariant energy scale,''
  Phys.\ Rev.\ D {\bf 67}, 044017 (2003)
  [arXiv:gr-qc/0207085].
  %%CITATION = GR-QC 0207085;%%

%\cite{Smolin:2005cz}
\bibitem{Smolin05cz}
  L.~Smolin,
  %``Falsifiable predictions from semiclassical quantum gravity,''
  Nucl.\ Phys.\ B {\bf 742}, 142 (2006)
  [arXiv:hep-th/0501091].
  %%CITATION = HEP-TH 0501091;%%

%\cite{Galan:2004st}
\bibitem{Galan04st}
  P.~Galan and G.~A.~Mena Marugan,
  %``Quantum time uncertainty in a gravity's rainbow formalism,''
  Phys.\ Rev.\ D {\bf 70}, 124003 (2004)
  [arXiv:gr-qc/0411089].
  %%CITATION = GR-QC 0411089;%%

%\cite{Galan:2005ju}
\bibitem{Galan05ju}
  P.~Galan and G.~A.~Mena Marugan,
  %``Length uncertainty in a gravity's rainbow formalism,''
  Phys.\ Rev.\ D {\bf 72}, 044019 (2005)
  [arXiv:gr-qc/0507098].
  %%CITATION = GR-QC 0507098;%%
%\cite{Hackett05mb}
\bibitem{Hackett05mb}
  J.~Hackett,
  %``Asymptotic flatness in rainbow gravity,''
  Class.\ Quant.\ Grav.\  {\bf 23}, 3833 (2006)
  [arXiv:gr-qc/0509103].
  %%CITATION = GR-QC 0509103;%%
%\cite{Aloisio:2005qt}
\bibitem{Aloisio05qt}
  R.~Aloisio, A.~Galante, A.~Grillo, S.~Liberati, E.~Luzio and F.~Mendez,
  % ``Deformed special relativity as an effective theory of measurements on
  %quantum gravitational backgrounds,''
  Phys.\ Rev.\ D {\bf 73}, 045020 (2006)
  [arXiv:gr-qc/0511031].
  %%CITATION = GR-QC 0511031;%%

%\cite{Ling:2005bp}
\bibitem{Ling05bp}
  Y.~Ling, X.~Li and H.~Zhang,
  ``Thermodynamics of modified black holes from gravity's rainbow,''
  [arXiv:gr-qc/0512084].
  %%CITATION = GR-QC 0512084;%%

%\cite{Galan:2006by}
\bibitem{Galan06by}
  P.~Galan and G.~A.~M.~Marugan,
  %``Entropy and temperature of black holes in a gravity's rainbow,''
  Phys.\ Rev.\ D {\bf 74}, 044035 (2006)
  [arXiv:gr-qc/0608061].
  %%CITATION = GR-QC 0608061;%%

\bibitem{Ling06}
  Y.Ling, ``Rainbow Universe''.
  [arXiv:gr-qc/0609129].
%\cite{Deser:1998bb}
\bibitem{Deser98bb}
  S.~Deser and O.~Levin,
  % ``Equivalence of Hawking and Unruh temperatures through flat space
  %embeddings,''
  Class.\ Quant.\ Grav.\  {\bf 15}, L85 (1998)
  [arXiv:hep-th/9806223].
  %%CITATION = HEP-TH 9806223;%%

%\cite{Deser:1998xb}
\bibitem{Deser98xb}
  S.~Deser and O.~Levin,
  %``Mapping Hawking into Unruh thermal properties,''
  Phys.\ Rev.\ D {\bf 59}, 064004 (1999)
  [arXiv:hep-th/9809159].
  %%CITATION = HEP-TH 9809159;%%

\end{thebibliography}
\end{document}